# Performance of Localized-Orbital Coupled Cluster Approaches for the Conformational Energies of Longer n-alkane Chains

Golokesh Santra* and Jan M.L. Martin*

Department of Molecular Chemistry and Materials Science, Weizmann Institute of Science,

7610001 Reḥovot, Israel.

Email: gershom@weizmann.ac.il; golokesh.santra@weizmann.ac.il

**Abstract.** We report an update and enhancement of the ACONFL (conformer energies of large alkanes [Ehlert, S.; Grimme, S.; Hansen, A. *J. Phys. Chem. A* **2022**, *126*, 3521-3535]) dataset. For the ACONF12 (n-dodecane) subset, we report basis set limit *canonical* CCSD(T) reference data obtained from MP2-F12/cc-pV{T,Q}Z-F12 extrapolation, [CCSD(F12*)-MP2-F12]/aug-cc-pVTZ-F12, and a (T) correction from conventional CCSD(T)/aug-cc-pV{D,T}Z calculations. Then we explored the performance of a variety of single and composite localized-orbital CCSD(T) approximations, ultimately finding an affordable LNO-CCSD(T)-based post-MP2 correction that agrees to 0.006 kcal/mol MAD (mean absolute deviation) with the revised canonical reference data. In tandem with canonical MP2-F12/CBS extrapolation, this was then used to re-evaluate the ACONF16 and ACONF20 subsets for *n*-hexadecane and *n*-icosane, respectively. Combining those with the revised canonical reference data for the dodecane conformers (i.e., ACONF12 subset) a revised ACONFL set was obtained. It was then used to assess the performance of different localized-orbital coupled cluster approaches, such as PNO-LCCSD(T) as implemented in MOLPRO, DLPNO-CCSD($T_0$) and DLPNO-CCSD($T_1$) as implemented in ORCA, and LNO-CCSD(T) as implemented in MRCC, at their respective "Normal", "Tight", "vTight" and "vvTight" accuracy settings. For a given accuracy threshold and basis set, DLPNO-CCSD($T_1$) and DLPNO-CCSD($T_0$) offer comparable performance. With "VeryTightPNO" cutoffs, explicitly correlated DLPNO-CCSD($T_1$)-F12/VDZ-F12 is the best pick among all the DLPNO-based methods tested. To isolate basis set incompleteness from localized orbital-related truncation errors (domain, localized natural orbitals), we have also compared the localized coupled cluster approaches with canonical DF-CCSD(T)/aug-cc-pVTZ for the ACONF12 set. We found that gradually tightening the cutoffs improves the performance of LNO-CCSD(T) and using a composite scheme such as vTight + 0.50[vTight − Tight] improves things further. For DLPNO-CCSD($T_1$), "TightPNO" and "VeryTightPNO" offers statistically similar accuracy, which gets slightly better when $T_{CutPNO}$ is extrapolated to the complete PNO space limit. Like Brauer *et al.* [*Phys. Chem. Chem. Phys.* **2016**, *18* (31), 20905–20925] previously reported for the S66x8 noncovalent interactions, the dispersion-corrected dRPA-based double hybrids perform remarkably well for the ACONFL set. While the revised reference data do not affect any conclusions on the less accurate methods, they may upend orderings for more accurate methods with error statistics on the same order as the difference between reference datasets.

I.     **Introduction.**

According to the IUPAC gold book,[1] a conformer is "*one of a set of stereoisomers (i.e., isomers that possess identical constitution, but which differ in the arrangement of their atoms in space[1]), each of which is characterized by a conformation corresponding to a distinct potential energy minimum.*" A comprehensive understanding of the conformers of organic and biomolecules is critical because they often show distinct chemical and biological activities.[2,3] Hence, accurate estimation of their structural properties and conformational energies (i.e., relative energies of other conformers with respect to the most stable one) is essential for a better understanding of biological phenomena, like protein folding, substrate binding, catalysis by enzymes, and many more (see refs.[4,5] and references therein). Multiple conformers of a certain molecule often span a relatively narrow conformational energy range. Hence, these structures exist as a thermally populated mixture at room or physiological temperature. Therefore, all relevant conformers must be considered to evaluate molecular properties accurately.[6]

Due to the high flexibility of open-chain molecules in conformational space,[7] these are often used for modeling flexible biological systems, e.g., in drug discovery and related applications.[8] One class of such systems are linear n-alkanes (i.e., $C_nH_{2n+2}$); simple as they may be, they constitute key building blocks of organic chemistry, of fossil fuels, and of polymers, lipids, and biomembranes.

Since the pioneering work of Pitzer,[9] the existence of different conformers of *n*-alkane is well known. Over the years, conformational enthalpies and low-lying conformers of shorter unbranched alkane chains have been studied experimentally.[10–12] On the other hand, theoretically, the torsional space and equilibrium conformational energies of short *n*-alkanes have also been explored theoretically.[13–17] Previous theoretical studies[18–21] on the longer *n*-alkane chains (n > 10) often investigated a handful of the lowest energy conformers to find the balance between repulsive hydrogen contacts and attractive London dispersion and finally predict the energy gap between the linear zigzag and hairpin-like conformers.

In 2009, Gruzman et al.[14] published comprehensive theoretical studies on conformational equilibrium energies of five unbranched n-alkanes isomers: *n*-butane, *n*-pentane, *n*-hexane, *n*-heptane, and *n*-octane (i.e., $C_nH_{2n+2}$; n=4-8). Later, to elucidate the correct treatment of dispersion, a detailed study of the entire conformer surface of n-pentane was performed by Martin.[13] Both experimental[19] and theoretical[22,23] studies have already established that due to the much stronger attractive London dispersion, the conformational preferences of longer unbranched n-alkanes (both in the gas and liquid phases) are very different from what we generally see for the short n-alkanes. Hence, a comprehensive study of longer n-alkane conformers beyond the n-octane is vital from chemical and biological perspective.

Recently, Ehlert et al.[24] have proposed a dataset, ACONFL, composed of the thirteen, seventeen, and twenty-one lowest-energy conformers of *n*-dodecane, *n*-hexadecane, and *n*-icosane, respectively. That is, the complete ACONFL set consists of 12 *n*-dodecane conformer energies (i.e., ACONF12), 16 *n*-hexadecane conformer energies (i.e., ACONF16), and 20 *n*-icosane conformer energies (i.e., ACONF20). In ref.[24], the reference conformer energies of ACONF12 were calculated at the DLPNO-CCSD($T_1$)/AV{T,Q}Z level with VeryTightPNO setting. As the CBS extrapolation for the ACONF16 and ACONF20 conformers are still too expensive, the authors in ref.[24] used the arithmetic mean of the δCBS and xCBS schemes. Both δCBS and xCBS are based on focal-point analysis. (For the details of δCBS and xCBS extrapolation techniques, see ref.[25] and [24], respectively.)

Although canonical CCSD(T) or explicitly correlated CCSD(T)-F12 are preferred for accurate conformer energies, due to the steep $N^7$ cost scaling of these methods with the system size, using them is often impractical for large molecules. Hence, linear-scaling localized coupled cluster methods such as the PNO-LCCSD(T) [pair natural orbital localized coupled-cluster with singles, doubles, and perturbative triples] method of Ma and Werner,[26] the DLPNO-CCSD(T) [domain localized pair natural orbital CCSD(T)] by Riplinger, Guo, Pinski, Valeev, and Neese,[27,28] and the LNO-CCSD(T) [localized natural orbital CCSD(T)] method of Nagy, Kallay, and coworkers[29–31] are attractive alternatives to canonical coupled cluster methods. With a sufficiently tight accuracy setting, these methods can achieve an accuracy similar to canonical CCSD(T) in the same basis set. Although the favorable linear cost scaling of the localized coupled cluster methods allows them to be used for systems with hundreds of atoms, their accuracy is subject to multiple predefined cutoffs. The fixed combinations for DLPNO-CCSD(T) are `LoosePNO`, `NormalPNO`, `TightPNO`, and `VeryTightPNO` (see Table 1 in Ref.[32] for definitions). The available accuracy thresholds for LNO-CCSD(T) are `Normal`, `Tight`, `vTight`, or `vvTight` (see Table 1 in Ref.[31] for details). In PNO-LCCSD(T), `Default` and `Tight` (see Tables 1-4 in Ref.[26] for more information) are the standard settings. Examples of recent use of these localized orbital coupled cluster methods include the energetics of the $(H_2O)_{20}$ cages using PNO-LCCSD(T)-F12b,[33] (the F12b suffix refers to explicit correlation[34]), the noncovalent interaction energies of seven large dimers (L7 set[35]) with LNO-CCSD(T),[36] the main group thermochemistry, barrier heights, intra-, and intermolecular interaction energies of GMTKN55[37] using DLPNO-CCSD(T),[38] and benchmark studies on the Ru(II) complexes involved in hydroarylation,[39] highly delocalized polypyrroles (POLYPYR[40] set), metal−organic barrier heights (MOBH35, 35 reactions[41–43]), an efficient estimation of formation enthalpies ($\Delta_f H^0$) for closed-shell organic compounds,[44] predicting gas-phase anion binding energies,[45] benchmarking of localized-orbital G4(MP2)-equivalents [46,47] composite WFT methods for fullerene isomerization energies,[48] etc.

Recently, in a conference proceeding extended abstract[49] and later in a full research article,[50] the present authors have evaluated the performance of pure and composite localized coupled cluster methods for the S66 and S66x8 noncovalent interaction sets, respectively. In Refs.[49,50] we found that LNO-CCSD(T) with a vvTight threshold can achieve canonical CCSD(T) level accuracy. Designing few low-cost composite methods, we obtained the noncovalent interaction energies close to the reference level.

Hence, the main objectives of the present study are: (a) improving the reference conformer energies of longer n-alkane chains and (b) evaluating different pure and composite localized orbital coupled cluster methods relative to the new reference conformer energies. As a by-product, some conclusions about DFT and other approximate methods can also be drawn.

## II. Computational Details.

Explicitly correlated CCSD(F12*), CCSD(T)-(F12*), PNO-LCCSD(T)-F12b, and regular PNO-LCCSD(T) single-point calculations were performed with Molpro.[51] The ORCA 5.0.3[52] package was employed for the RI-MP2, explicitly correlated canonical RI-MP2-F12[53] with *ansatz* 3C(FIX), DLPNO-CCSD($T_1$)-F12, and DLPNO-CCSD($T_1$) calculations. Finally, for the density fitted canonical CCSD(T) [i.e., DF-CCSD(T)] and for LNO-CCSD(T), we used MRCC 2022.[54] All the calculations were carried out at the Faculty of Chemistry HPC facility at the Weizmann Institute of Science.

For the RI-MP2-F12 and CCSD(F12*), we employed the cc-pVnZ-F12[55] (n=T, Q) and cc-pVTZ-F12[55] orbital basis set, respectively. Suitable OptRI, JKfit,[56] (for Coulomb and exchange) and MP2fit[57,58] (density fitting in MP2) basis set combinations were used throughout the F12 calculations. The DF-CCSD(T) calculations were carried out using correlation consistent aug-cc-pV$n$Z (n = D, T)[59,60] orbital basis sets and the corresponding RI fitting[57,58] basis sets.

For PNO-LCCSD(T)-F12b and DLPNO-CCSD($T_1$)-F12 calculations, we have used cc-pVnZ-F12 (n=D, T) basis set, with suitable JKfit[56] and CABS (complementary auxiliary basis set[61]) options. Geminal Slater exponent (β) values of 0.9 and 1.0 were used for the pp-pVDZ-F12 and cc-pVTZ-F12, respectively. All localized orbital F12 calculations were performed using density fitting (DF).

Correlation consistent aug-cc-pVnZ (n = T, Q, and 5) basis sets were used for the localized orbital coupled cluster calculations, together with suitable JKfit basis sets for the Coulomb and exchange energy and RI fitting basis sets aug-cc-pVnZ-RI[57,58] (n = T, Q, and 5) for the correlation component. For LNO-CCSD(T), we have used Normal, Tight, vTight, and vvTight accuracy thresholds. On the other hand, Default and Tight settings were employed for PNO-LCCSD(T). Following a suggestion by Prof. H.-J. Werner (personal conversation with the senior author), we have used the MOLPRO distance criterion `REXT=0` for all PNO-LCCSD(T) and PNO-LCCSD(T)-F12b calculations. [The default `REXT` setting for the "Tight" and "Default" domains are 7 and 5 a.u., respectively. By using `REXT=0`, the PAO domains are selected based solely on the connectivity criterion `IEXT` (2 and 3, for the "Tight" and "Default" settings, respectively) only. For the PNO-LCCSD(T) and PNO-LCCSD(T)-F12b statistics with the default "`REXT`" values, see Tables S8 and S9 in the Supporting Information.]

For DLPNO-CCSD($T_1$), we have used NormalPNO, TightPNO, and VeryTightPNO thresholds together with RIJCOSX (resolution of the identity in combination with the chain of spheres[62] algorithm) approximation. To investigate the dependence of the DLPNO-CCSD($T_1$) correlation on the size of the PNO space, we have considered two $T_{CutPNO}$ (the occupation number cutoff for a PNO to be included for a given electron pair) values ($10^{-X}$; X = 6 and 7) with TightPNO threshold. Two-point PNO extrapolations were also carried out to the complete PNOs space limit (CPS), using the simple two-point extrapolation scheme proposed by Altun $et\ al.$,[63] $E = E^X + [Y^\beta/(Y^\beta - X^\beta)] \times (E^Y - E^X)$, where Y=X+1 and β=7.13. This corresponds numerically to $E^X + 1.5 \times (E^Y - E^X)$, or equivalently and perhaps more clearly $E^Y + 0.5 \times (E^Y - E^X)$.

For two-point complete basis set extrapolation, we have employed the expression from Halkier *et al.*[64], $E_{CBS} = E_L + (E_L - E_{L-1})/[\left(\frac{L}{L-1}\right)^\alpha - 1]$, where L refers to the basis set cardinal number, and α is the basis set extrapolation exponent. Following Hill *et al.*[65], we have used α=4.355 and 2.531 for the RI-MP2-F12/V{T,Q}Z-F12 and RI-MP2/AV{T,Q}Z energies, respectively. While extrapolating the RI-MP2-F12 energies to the CBS limit, the SCF component was taken from the largest basis set calculation with CABS[61] (complementary auxiliary basis set) correction and only MP2-F12 components were extrapolated. For the AV{D,T}Z extrapolation of the canonical perturbative triples term, we have used α=3.096, as recommended by Schwenke.[66] Like the W1 and W2 theories,[67] for the localized coupled cluster methods, we used the extrapolation exponents 3.22 and 3.0 for the AV{T,Q}Z and AV{Q,5}Z extrapolations, respectively.

Most of the DFT, semiempirical quantum mechanical, and force field (FF) results were extracted from the ESI of Ref.[24] The revDSD-PBEP86-D3BJ,[68] revDSD-PBEP86-D4,[68] and ωB97M(2)[69] functionals were evaluated using Q-CHEM 6.[70] On the other hand, dRPA75[71] and DSD-dRPA[72] calculations were performed using MRCC 2022.[54]

Geometries of all the conformers of *n*-dodecane ($C_{12}H_{26}$), *n*-hexadecane ($C_{16}H_{34}$), and *n*-icosane ($C_{20}H_{42}$) were extracted from ref.[24]

**III.   Results and Discussion.**

*(a) Revised reference conformer energies of n-dodecane:*

As the first step, we calculated the canonical explicitly correlated RI-MP2-F12 energies of the *n*-dodecane conformers with cc-pVTZ-F12 and cc-pVQZ-F12 basis sets and extrapolated them to the CBS limit to eliminate the basis set incompleteness error (BSIE). On top of that, we have used the [CCSD(F12*) − MP2-F12]/cc-pVTZ-F12 energies for the CCSD-MP2 term of the high-level correction (HLC), while the perturbative triples [i.e., (T)] contribution is taken from the DF-CCSD(T)/AV{D,T}Z calculation. Hence, the final equation for the calculation of the reference conformer energies of *n*-dodecane is:

$$\Delta E_{conf} = \Delta E_{MP2-F12/V\{T,Q\}Z-F12} + \Delta\Delta E_{[CCSD(F12*)-MP2-F12]/VTZ-F12}$$
$$+ \Delta\Delta E_{[DF-CCSD(T)-DF-CCSD]/AV\{D,T\}Z}$$

Based on the HLCs used for the S66 non-covalent interaction energies, Kesharwani et al.[73], inspired by an earlier study by Sherrill and coworkers,[74] proposed a hierarchy of noncovalent interaction energies: *gold* (employing [CCSD(F12*)–MP2-F12]/VQZ-F12 + (T)/haV{T,Q}Z), *silver* (using [CCSD(F12*)–MP2-F12]/VTZ-F12 + (T)/haV{D,T}Z), and *bronze* (employing CCSD(F12*)(Tc$_{sc}$)/cc-pVDZ-F12). Hence, the revised reference data for the ACONF12 set can be considered a "silver" standard.

**Table 1:** Mean absolute deviations (MADs, kcal/mol), root-mean-square deviations (RMSDs, kcal/mol), and mean signed deviations (MSDs, kcal/mol) of explicitly correlated CCSD(F12*)/VTZ-F12, canonical DF-CCSD(T)/AVnZ, and DLPNO-CCSD(T$_1$,VeryTightPNO)/CBS (i.e., the reference conformer energies reported in ref[24]) for the ACONF12 set. Five CCSD(T)-F12*/VTZ-F12 level energies calculated in the present study have also been included.

| # Conf. (n=12) | ΔE$_{conf}$ (kcal/mol) | | | | | | |
|---|---|---|---|---|---|---|---|
| | MP2-F12 /V{T,Q}Z-F12 +HLC(CCSD)[a] +HLC(T)[b] | DLPNO-CCSD(T$_1$, VeryTightPNO) /CBS[c] (From ref[24]) | CCSD(F12*) /VTZ-F12 | DF-CCSD(T) /AVDZ | DF-CCSD(T) /AVTZ | DF-CCSD(T) /AV{D,T}Z[d] | CCSD(T)-(F12*) /VTZ-F12[e] |
| 1 | 1.82 | 1.96 | 2.09 | 1.64 | 1.66 | 1.63 | 1.84 |
| 2 | 2.05 | 2.27 | 2.57 | 1.46 | 1.70 | 1.76 | 2.08 |
| 3 | 2.49 | 2.70 | 2.96 | 2.00 | 2.19 | 2.22 | 2.52 |
| 4 | 3.16 | 3.48 | 3.93 | 1.89 | 2.47 | 2.71 | 3.18 |
| 5 | 3.66 | 3.85 | 4.08 | 3.17 | 3.34 | 3.39 | — |
| 6 | 3.88 | 4.09 | 4.47 | 3.03 | 3.41 | 3.56 | n/a |
| 7 | 4.16 | 4.39 | 4.67 | 3.59 | 3.78 | 3.82 | 4.19 |
| 8 | 4.31 | 4.62 | 5.11 | 3.15 | 3.64 | 3.84 | — |
| 9 | 4.89 | 5.16 | 5.53 | 4.19 | 4.46 | 4.54 | — |
| 10 | 5.45 | 5.77 | 6.24 | 4.15 | 4.79 | 5.06 | — |
| 11 | 5.99 | 6.30 | 6.78 | 4.96 | 5.38 | 5.55 | — |
| 12 | 6.56 | 6.80 | 7.22 | 6.03 | 6.14 | 6.15 | — |
| **MAD** | Reference | **0.25** | **0.60** | **0.76** | **0.46** | **0.35** | |
| **RMSD** | | **0.25** | **0.62** | **0.84** | **0.48** | **0.36** | |
| **MSD** | | 0.25 | 0.60 | -0.76 | -0.46 | -0.35 | |

[a] HLC(CCSD)=[CCSD(F12*) − MP2-F12]/VTZ-F12
[b] HLC(T)=[DF-CCSD(T) − DF-CCSD]/AV{D,T}Z
[c] VeryTightPNO setting according to ref,[24] i.e., TightPNO with T$_{CutPNO}$=10$^{-8}$, T$_{CutMKN}$=10$^{-4}$, & T$_{CutPairs}$=10$^{-6}$.
[d] SCF, CCSD, and (T) components are extrapolated separately, using Schwenke's exponents α=3.342, 2.451, and 3.096, respectively.[66]
[e] Following Ref.[75], the (T) term of CCSD(T)-(F12*) is scaled by 1.0527.

Ehlert et al.'s DLPNO-CCSD(T$_1$)/CBS energies evaluated with a VeryTightPNO threshold have MAD=0.25 kcal/mol relative to the presently revised reference data. The mean absolute error for the density fitted canonical CCSD(T) with two-point CBS extrapolation from the aug-cc-pVnZ (n=D and T) energies is 0.35 kcal/mol [see footnote (d) of Table 1 for the extrapolation details]. A detailed inspection of the signed deviations for the twelve *n*-dodecane conformer energies reveals that DLPNO-CCSD($T_1$)/CBS overestimates, but DF-CCSD(T) underestimates, the conformer energies across the board (see Table 1).

Owing to their CPU time and resource requirements (over 3 weeks wall clock each, 512GB of RAM, and 3.3 TB of SSD scratch disk), we were only able to calculate five CCSD(T)-(F12*)/VTZ-F12 level conformer energies. We note that unlike CCSD, F12 approaches do not benefit the connected quasiperturbative triples,[76] so the basis set convergence behavior of the (T) contribution is effectively the same as what one observes for the conventional CCSD(T) calculations. In ref.[75], Peterson *et al.* proposed a global scale factor, 1.0527, for scaling the (T) component of the CCSD(T)-F12b/VTZ-F12 level atomization energies. In the present study, we have applied this so-called ($T_s$) approximation in the CCSD($T_s$)(F12*)/VTZ-F12 calculations. Table 1 shows that the reference-level *n*-dodecane conformer energies are very close to those obtained using (T)-scaled CCSD(T)-(F12*)/VTZ-F12.

Now, let us take a closer look at the performance of RI-MP2 and explicitly correlated RI-MP2-F12 methods. Even with the cc-pVDZ-F12 basis set, RI-MP2-F12 achieves the complete basis set limit (MAD=0.01 kcal/mol, with respect to the RI-MP2-F12/AV{T,Q}Z-F12 energies; see Table S1 in the Supporting Information). However, if we assess the performance of RI-MP2 and RI-MP2-F12 against the "new" reference data of *n*-dodecane conformers, we found that RI-MP2/AV{T,Q}Z is the best performer (MAD=0.18 kcal/mol). Considering the fact that the mean absolute difference between RI-MP2/AV{T,Q}Z and RI-MP2-F12/AV{T,Q}Z-F12 is 0.15 kcal/mol, we can safely say that the former method gets a better answer for the wrong reasons. Canonical RI-MP2 and explicitly correlated RI-MP2-F12 systematically underestimate the conformer energies with respect to the "new" revised reference (see Table S1 in the Supporting Information). Interestingly, with {T,Q}-extrapolation the accuracy of RI-MP2-F12 (MAD=0.33 kcal/mol) is only marginally worse than DLPNO-CCSD($T_1$)/VeryTightPNO (MAD=0.25 kcal/mol).

Next, in order to eliminate basis set incompleteness as a 'confounding factor', we use the aug-cc-pVTZ basis set throughout to assess the performance for ACONF12 of LNO-CCSD(T), PNO-LCCSD(T), and DLPNO-CCSD($T_1$) relative to canonical DF-CCSD(T) (see Table 2). As expected, tightening the accuracy threshold for LNO-CCSD(T) improves its accuracy. Proposed by Nagy et al.[31], the low-cost composite method Tight + 0.5[Tight − Normal] performs similarly to LNO-CCSD(T, vTight). With a mean absolute error of 0.01 kcal/mol, vTight + 0.5[vTight − Tight] offers the best accuracy among all the individual, and composite LNO-CCSD(T) tested (see Table 2). Instead of using a fixed 0.5 prefactor in the composite schemes, when we optimized the prefactors with respect to canonical CCSD(T)/AVTZ conformer energies, we obtained slightly different values: 0.84 for Tight + A[Tight − Normal] and 0.60 for vTight + A[vTight − Tight] (see Table 2). With a MAD of 0.03 kcal/mol, Tight + 0.84[Tight − Normal] is closer to the accuracy of vTight + 0.5[vTight − Tight] than Tight + 0.5[Tight − Normal]. From the MSD values listed in Table 2, it is clear that standard and composite LNO-CCSD(T) methods underestimate the conformer energies relative to canonical DF-CCSD(T).

Standard PNO-LCCSD(T) with Default and Tight settings significantly overestimate the ACONF12 energies (see Table 2). The PNO-based composite scheme, Tight + 0.50[Tight – Default], performs only marginally better than the standard alternatives. Upon optimization, the composite method, Tight + 3.32[Tight – Default], offers significantly better accuracy, but with an anomalously large coefficient.

As expected, standard and composite DLPNO-CCSD($T_1$) is marginally better than DLPNO-CCSD($T_0$) for any given accuracy setting. DLPNO-CCSD($T_1$)/TightPNO with CPS extrapolation from $T_{CutTNO}=\{10^{-6}, 10^{-7}\}$ performs better than DLPNO-CCSD($T_1$)/VeryTightPNO. The latter method is clearly a better performer than the composite method, TightPNO + 0.5[TightPNO – NormalPNO], albeit at ca. 4 times the computational cost.

As DLPNO-CCSD($T_1$) is much more demanding in terms of I/O, storage, and bandwidth requirements than DLPNO-CCSD($T_0$),[39,41,42] the DLPNO-CCSD($T_0$)/TightPNO/CPS can be a more economical alternative to DLPNO-CCSD($T_1$)/TightPNO/CPS.

Refitting of the DLPNO-based composite methods with respect to DF-CCSD(T) level ACONF12 energies offered significantly better accuracy but with unphysical negative coefficients (see Table 2).

**Table 2:** Mean absolute deviations (MADs, kcal/mol), root-mean-square deviations (RMSDs, kcal/mol), and mean signed deviations (MSDs, kcal/mol) of different standard and composite localized coupled cluster methods relative to the canonical DF-CCSD(T) level conformer energies of *n*-dodecane. We have used the aug-cc-pVTZ basis set throughout. Heatmapping is from red (worst) via yellow to green (best).

| Methods | Threshold | Coeff. (A) | $T_{CutPNO}$ | MAD (kcal/mol) | RMSD (kcal/mol) | MSD (kcal/mol) |
|---|---|---|---|---|---|---|
| LNO-CCSD(T) | Normal | | | 0.31 | 0.32 | -0.31 |
| | Tight | | | 0.15 | 0.15 | -0.15 |
| | vTight | | | 0.05 | 0.06 | -0.05 |
| | vvTight | | | 0.05 | 0.05 | -0.05 |
| | Tight + A[Tight − Normal][a] | 0.50 | | 0.07 | 0.08 | -0.07 |
| | vTight + A[vTight − Tight][a] | 0.50 | | 0.01 | 0.02 | -0.01 |
| | Tight + A[Tight − Normal] | 0.84 | | 0.03 | 0.04 | -0.01 |
| | vTight + A[vTight − Tight] | 0.60 | | 0.01 | 0.01 | 0.00 |
| PNO-LCCSD(T) | Default | | | 0.49 | 0.51 | 0.49 |
| | Tight | | | 0.38 | 0.40 | 0.38 |
| | Tight + A[Tight − Default] | 0.50 | | 0.33 | 0.35 | 0.33 |
| | Tight + A[Tight − Default] | 3.32 | | 0.05 | 0.08 | 0.02 |
| DLPNO-CCSD(T$_0$) | NormalPNO | | | 0.22 | 0.23 | -0.22 |
| | TightPNO | | $T_{CutPNO}=10^{-6}$ | 0.16 | 0.16 | 0.16 |
| | | | $T_{CutPNO}=10^{-7}$ | 0.07 | 0.08 | 0.07 |
| | | | CPS[b] or 1.0x $10^{-\{6,7\}}$ | 0.03 | 0.03 | 0.03 |
| | VeryTightPNO | | | 0.09 | 0.10 | 0.09 |
| | TightPNO + A[TightPNO − NormalPNO] | 0.50 | | 0.22 | 0.22 | 0.22 |
| | TightPNO + A[TightPNO − NormalPNO] | -0.24 | | 0.01 | 0.02 | 0.00 |
| DLPNO-CCSD(T$_1$) | NormalPNO | | | 0.24 | 0.25 | -0.24 |
| | TightPNO | | $T_{CutPNO}=10^{-6}$ | 0.13 | 0.14 | 0.13 |
| | | | $T_{CutPNO}=10^{-7}$ | 0.05 | 0.05 | 0.05 |
| | | | CPS[b] or 1.0x $10^{-\{6,7\}}$ | 0.01 | 0.02 | 0.00 |
| | VeryTightPNO | | | 0.07 | 0.08 | 0.07 |
| | TightPNO + A[TightPNO − NormalPNO] | 0.50 | | 0.19 | 0.19 | 0.19 |
| | TightPNO + A[TightPNO − NormalPNO] | -0.15 | | 0.01 | 0.02 | 0.00 |

[a]Composite methods proposed in Ref.[31]
[b] CPS=complete PNO space limit; $E_{CPS}=E^X + [Y^\beta/(Y^\beta − X^\beta)]*(E^Y − E^X)$, where Y=X+1 and β=7.13, which corresponds numerically to $E_{CPS}=E^Y + 0.5(E^Y − E^X)$ (see Ref.[63])

*(b) Revised reference conformer energies of n-hexadecane and n-icosane conformers (i.e., the ACONF16 and ACONF20 subsets):*

In the nomenclature of Hansen and coworkers,[24] the conformer **0** is "all trans" and **00** is "hairpin"-like (see Figures 2-3 in ref.[24] for illustration). As the chain grows longer, eventually dispersion forces will favor a folded over a linear structure. The "all trans" conformer is the lowest in energy for *n*-dodecane, while the "hairpin" is clearly the global minimum for *n*-icosane; *n*-hexadecane lies near the transition point.[19]

**Table 3:** List of localized orbital based high-level corrections and their performance relative to the canonical HLC used on top of the RI-MP2-F12/CBS level conformer energies for the revised reference conformer energies of the ACONF12 set. All the results are in kcal/mol.[a]

| HLCs | Acronyms | MAD (kcal/mol) | RMSD (kcal/mol) | MSD (kcal/mol) |
|---|---|---|---|---|
| [CCSD(F12*) − MP2-F12]/VTZ-F12 + (T)/AV{D,T}Z | | Reference | | |
| [DF-CCSD(T) − RI-MP2]/AVTZ | HLC1 | 0.035 | 0.038 | 0.035 |
| [DLPNO-CCSD(T$_1$) − LMP2]/TightPNO/CPS{6,7}/AVTZ | HLC2 | 0.050 | 0.053 | 0.050 |
| [DLPNO-CCSD(T$_1$) − LMP2]/VeryTightPNO/AVTZ | HLC3 | 0.073 | 0.078 | 0.073 |
| [DLPNO-CCSD(T$_1$) − LMP2]/ VeryTightPNO/AV{T,Q}Z | HLC4 | 0.125 | 0.132 | 0.125 |
| [DLPNO-CCSD(T$_1$) − LMP2]/TightPNO/CPS{6,7}/AV{T,Q}Z | HLC5 | 0.130 | 0.132 | 0.130 |
| [LNO-CCSD(T) − LMP2]/Tight/AVTZ | HLC6 | 0.032 | 0.034 | -0.032 |
| [LNO-CCSD(T) − LMP2]/vTight/AVTZ | HLC7 | 0.017 | 0.021 | 0.016 |
| [LNO-CCSD(T) − LMP2]/vvTight/AVTZ | HLC8 | 0.016 | 0.023 | 0.015 |
| [LNO-CCSD(T) − LMP2]/Tight/AVQZ | HLC9 | 0.009 | 0.011 | -0.007 |
| [LNO-CCSD(T) − LMP2]/vTight/AVQZ | HLC10 | 0.059 | 0.062 | 0.059 |
| [LNO-CCSD(T) − LMP2]/vTight/AV5Z | HLC11 | 0.031 | 0.032 | 0.031 |
| [LNO-CCSD(T) − LMP2]/vTight/AV{T,Q}Z | HLC12 | 0.088 | 0.091 | 0.088 |
| [LNO-CCSD(T) − LMP2]/Tight/AV{T,Q}Z | HLC13 | 0.012 | 0.014 | 0.010 |
| **[LNO-CCSD(T) − LMP2]/vTight/AV{Q,5}Z** | **HLC14** | **0.006** | **0.008** | **0.000** |
| [PNO-LCCSD(T) − LMP2]/Default/AVTZ | HLC15 | 0.015 | 0.018 | 0.015 |
| [PNO-LCCSD(T) − LMP2]/Tight/AVTZ | HLC16 | 0.038 | 0.039 | 0.038 |
| [PNO-CCSD(T) − LMP2]/Default/AVQZ | HLC17 | 0.035 | 0.035 | 0.035 |
| [PNO-CCSD(T) − LMP2]/Tight/AVQZ | HLC18 | 0.052 | 0.053 | 0.052 |
| [PNO-LCCSD(T) − LMP2]/Default/AV{T,Q}Z | HLC19 | 0.048 | 0.048 | 0.048 |
| [PNO-LCCSD(T) − LMP2]/Tight/AV{T,Q}Z | HLC20 | 0.062 | 0.063 | 0.062 |
| [PNO-LCCSD(Ts)-F12b − LMP2-F12]/ Default/VTZ-F12 | HLC21 | 0.087 | 0.089 | 0.087 |
| [PNO-LCCSD(Ts)-F12b − LMP2-F12]/Tight/VTZ-F12 | HLC22 | 0.091 | 0.091 | 0.091 |

[a] CPS=complete PNO space limit; $E_{CPS}=E^X + [Y^β/( Y^β − X^β)]*(E^Y − E^X)$, where Y=X+1 and β=7.13. (see Ref.[63]) The expression CPS{X,Y} refers to the extrapolation of $T_{CutPNO}$ to the CPS limit using $T_{CutPNO}=10^{-X}$ and $10^{-Y}$.

The high-level correction (HLC) we have used to calculate the revised *n*-dodecane conformer energies is even for them computationally quite expensive, and would become intractable for the larger species. Hence, linear scaling localized coupled-cluster methods would be attractive alternatives for HLC on top of the RI-MP2-F12/CBS level *n*-hexadecane and *n*-icosane conformer energies. The present study considers 22 such alternative HLCs (see Table 3). The LNO-based high-level correction, [LNO-CCSD(T) − LMP2]/vTight/AV{Q,5}Z is the best in class, very closely followed by five other options, HLC7, HLC8, HLC9, HLC13, and HLC15.

Among these five alternative HLCs, the remarkable accuracy of HLC9, HLC7, and HLC15 can be attributed to fortuitous error compensation between [LCCSD-LMP2] and (T) contributions. Two low-cost alternatives to HLC14 are HLC13 and HLC8. With CBS extrapolation, DLPNO-CCSD($T_1$)-based high-level corrections (i.e., HLC4 and HLC5) have the largest deviations relative to the canonical reference data.

**Table 4:** Our best estimates of the ACONFL conformer energies. For convenience, we have retained the numbering and ordering of different conformers from ref.[24] ; conformer energies of *n*-dodecane (i.e., ACONF12 set) and *n*-hexadecane (i.e., ACONF16 set) are relative to the all-trans conformer **0**, while those for *n*-icosane (i.e., ACONF20 set) are relative to the "hairpin" conformer **00**.

| # Conformers ACONF12 | $\Delta E_{conf}$ (kcal/mol)[a] | # Conformers ACONF16 | $\Delta E_{conf}$ (kcal/mol)[b] | # Conformers ACONF20 | $\Delta E_{conf}$ (kcal/mol)[b] |
|---|---|---|---|---|---|
| 1 | 1.82 | 00 | -0.49 | 0 | 2.20 |
| 2 | 2.05 | 1 | 2.15 | 1 | 4.15 |
| 3 | 2.49 | 3 | 2.54 | 5 | 4.81 |
| 4 | 3.16 | 4 | 2.68 | 6 | 4.99 |
| 5 | 3.66 | 2 | 2.94 | 7 | 5.27 |
| 6 | 3.88 | 6 | 3.24 | 3 | 4.85 |
| 7 | 4.16 | 7 | 3.28 | 11 | 5.60 |
| 8 | 4.31 | 5 | 3.34 | 10 | 5.58 |
| 9 | 4.89 | 8 | 3.68 | 4 | 5.31 |
| 10 | 5.45 | 9 | 3.98 | 12 | 5.74 |
| 11 | 5.99 | 10 | 4.08 | 8 | 5.28 |
| 12 | 6.56 | 11 | 4.35 | 2 | 5.05 |
| | | 12 | 4.54 | 9 | 5.77 |
| | | 14 | 4.95 | 16 | 6.01 |
| | | 13 | 5.02 | 13 | 6.12 |
| | | 15 | 5.84 | 17 | 6.33 |
| | | 16 | 6.11 | 15 | 6.52 |
| | | | | 19 | 6.65 |
| | | | | 14 | 6.38 |
| | | | | 18 | 6.74 |
| | | | | 20 | 7.94 |

[a] $\Delta E_{conf} = \Delta E_{MP2-F12/V\{T,Q\}Z-F12} + \Delta\Delta E_{[CCSD(F12^*) - MP2-F12]/VTZ-F12} + \Delta\Delta E_{(T)/AV\{D,T\}Z}$

[b] $\Delta E_{conf} = \Delta E_{MP2-F12/V\{T,Q\}Z-F12} + \Delta\Delta E_{[LNO-CCSD(T) - LMP2]/vTight/AV\{Q,5\}Z}$

A number of the HLCs only involve AVTZ basis sets, hence they may be directly compared with canonical [DF-CCSD(T) − RI-MP2]/AVTZ (i.e., HLC1). We found that HLC16 has only 0.009 kcal/mol mean absolute error, which is due to a substantial error compensation between [CCSD-MP2] and (T) contributions (see Table S2 in the Supporting information).

We have finally selected [RI-MP2-F12/CBS + HLC14] for the revised reference conformer energies of $n$-hexadecane and $n$-icosane. (For the high-level correction energies of individual species, see Table S3 in the Supporting Information.)

Our best estimates of the ACONFL conformer energies are listed in Table 4. To summarize: the conformer energies of $n$-dodecane are evaluated canonically using MP2-F12/cc-pV{T,Q}Z-F12 + [CCSD(F12*) − MP2-F12]/cc-pVTZ-F12 + (T)/aug-cc-pV{D,T}Z, while for the $n$-hexadecane and $n$-icosane conformers we have employed canonical MP2-F12/cc-pV{T,Q}Z-F12 + localized [LNO-CCSD(T) − LMP2]/vTight/aug-ccPV{Q,5}Z. Differences between the revised reference data and the original ACONFL reference conformer energies can reach as positive as +0.60 kcal/mol and as negative as −0.67 kcal/mol; moreover, for multiple $n$-icosane conformers the energetic ordering is upended.

For the ACONFL conformer energies employing HLC13 and HLC8, see Table S4 in the Supporting Information. [See Table S5 in the Supporting Information for the ACONF16 conformer energies relative to the **00** (hairpin) conformer.]

*(c) Performance of localized orbital coupled cluster methods:*

In this section, we assess the performance of LNO-CCSD(T), PNO-LCCSD(T), and DLPNO-CCSD($T_0$), and DLPNO-CCSD($T_1$) methods in combination with different accuracy thresholds and basis sets. Table 5 summarizes the mean absolute deviations (MADs), mean signed deviations (MSDs), and root-mean-square deviations (RMSDs) of different methods.

Increasing the basis set size for a given accuracy threshold improves LNO-CCSD(T) accuracy relative to our revised reference data. As long as only single basis sets are considered, with MAD=0.06 kcal/mol, LNO-CCSD(T,vTight)/AV5Z is the best pick. Except for the "Normal" threshold, AV{Q,5}Z extrapolated results have lower mean absolute error compared to the respective AV{T,Q}Z energies. The excellent performance of LNO-CCSD(T,Normal)/AV{T,Q}Z (MAD=0.05 kcal/mol) is due to fortunate error compensation between localized orbital (LO) error and basis set incompleteness error. Increasing the chain length of n-alkane also increases the mean absolute deviation from the reference conformer energies accordingly (see Table 5). With a mean absolute error of 0.04 kcal/mol, LNO-CCSD(T,Tight)/AV{Q,5}Z is the best pick among the standard LNO-CCSD(T) methods tested with different basis sets and accuracy thresholds. However, if we consider the statistical uncertainty of the reference energies, the performance of LNO-CCSD(T,Tight)/AV{Q,5}Z, LNO-CCSD(T,vTight)/AV{Q,5}Z, and LNO-CCSD(T,vTight)/AV5Z are actually indistinguishable. While employing the AVTZ basis set for calculation, tightening the accuracy cutoffs from "vTight" to "vvTight" has no additional advantage.

**Table 5:** Performance of standard and composite LNO-CCSD(T), PNO-LCCSD(T), DLPNO-CCSD($T_0$), and DLPNO-CCSD($T_1$) methods with respect to the revised ACONFL reference data. Heatmapping is from red (worst) via yellow to green (best).[a]

| Method Details | Threshold | Basis set | MAD (kcal/mol) | | | | MSD (kcal/mol) | RMSD (kcal/mol) |
|---|---|---|---|---|---|---|---|---|
| | | | ACONFL | ACONF12 | ACONF16 | ACONF20 | | |
| LNO-CCSD(T) | Normal | AVTZ | 0.95 | 0.77 | 0.98 | 1.03 | -0.13 | 1.05 |
| | | AVQZ | 0.36 | 0.29 | 0.38 | 0.38 | -0.06 | 0.39 |
| | | AV5Z | 0.26 | 0.21 | 0.22 | 0.31 | 0.00 | 0.28 |
| | | AV{T,Q}Z | 0.05 | 0.03 | 0.03 | 0.09 | -0.01 | 0.07 |
| | | AV{Q,5}Z | 0.17 | 0.14 | 0.09 | 0.25 | 0.06 | 0.20 |
| | Tight | AVTZ | 0.80 | 0.60 | 0.80 | 0.93 | -0.08 | 0.91 |
| | | AVQZ | 0.22 | 0.16 | 0.22 | 0.25 | -0.03 | 0.25 |
| | | AV5Z | 0.10 | 0.08 | 0.12 | 0.10 | -0.03 | 0.12 |
| | | AV{T,Q}Z | 0.17 | 0.13 | 0.16 | 0.19 | 0.01 | 0.19 |
| | | AV{Q,5}Z | 0.04 | 0.01 | 0.03 | 0.08 | -0.04 | 0.06 |
| | vTight | AVTZ | 0.73 | 0.51 | 0.69 | 0.89 | -0.03 | 0.83 |
| | | AVQZ | 0.17 | 0.11 | 0.14 | 0.23 | 0.01 | 0.20 |
| | | AV5Z | 0.06 | 0.04 | 0.04 | 0.08 | 0.01 | 0.07 |
| | | AV{T,Q}Z | 0.19 | 0.15 | 0.22 | 0.20 | 0.04 | 0.22 |
| | | AV{Q,5}Z | 0.07 | 0.05 | 0.06 | 0.09 | 0.00 | 0.08 |
| | vvTight | AVTZ | 0.72 | 0.50 | 0.69 | 0.87 | -0.03 | 0.82 |
| PNO-LCCSD(T) | Default | AVTZ | 0.05 | 0.04 | 0.04 | 0.07 | -0.01 | 0.06 |
| | | AVQZ | 0.11 | 0.12 | 0.16 | 0.07 | 0.07 | 0.12 |
| | | AV5Z | — | 0.10 | 0.18 | — | — | — |
| | | AV{T,Q}Z | 0.17 | 0.18 | 0.24 | 0.12 | 0.11 | 0.19 |
| | | AV{Q,5}Z | — | 0.08 | 0.21 | — | — | — |
| | Tight | AVTZ | 0.08 | 0.07 | 0.09 | 0.07 | -0.05 | 0.09 |
| | | AVQZ | 0.09 | 0.11 | 0.13 | 0.05 | 0.06 | 0.10 |
| | | AV5Z | — | 0.10 | — | — | — | — |
| | | AV{T,Q}Z | 0.20 | 0.23 | 0.29 | 0.12 | 0.14 | 0.23 |
| | | AV{Q,5}Z | — | 0.10 | — | — | — | — |
| DLPNO-CCSD($T_0$) | NormalPNO | AVTZ | 0.73 | 0.68 | 0.88 | 0.63 | -0.24 | 0.79 |
| | | AVQZ | 0.28 | 0.32 | 0.39 | 0.17 | -0.16 | 0.32 |
| | | AV5Z | 0.22 | 0.28 | 0.33 | 0.10 | -0.19 | 0.26 |
| | | AV{T,Q} | 0.12 | 0.09 | 0.07 | 0.17 | -0.11 | 0.14 |
| | | AV{Q,5} | 0.22 | 0.23 | 0.27 | 0.17 | -0.21 | 0.23 |

| | | AVTZ | 0.51 | 0.30 | 0.43 | 0.69 | 0.05 | 0.60 |
|---|---|---|---|---|---|---|---|---|
| | TightPNO $T_{CutPNO}=10^{-6}$ | AVQZ | 0.08 | 0.10 | 0.08 | 0.08 | -0.04 | 0.10 |
| | | AV5Z | — | 0.16 | — | — | — | — |
| | | AV{T,Q} | 0.21 | 0.05 | 0.14 | 0.36 | -0.09 | 0.28 |
| | | AV{Q,5} | — | 0.42 | — | — | — | — |
| | TightPNO (Default or $T_{CutPNO}=10^{-7}$) | AVTZ | 0.62 | 0.38 | 0.54 | 0.82 | 0.03 | 0.73 |
| | | AVQZ | 0.14 | 0.04 | 0.06 | 0.25 | 0.08 | 0.19 |
| | | AV5Z | — | 0.05 | — | — | — | — |
| | | AV{T,Q} | 0.21 | 0.23 | 0.31 | 0.13 | 0.12 | 0.24 |
| | | AV{Q,5} | — | 0.11 | — | — | — | — |
| | TightPNO $T_{CutPNO}=10^{-\{6,7\}}$ or CPS{6,7}[a] | AVTZ | 0.68 | 0.42 | 0.60 | 0.89 | 0.03 | 0.79 |
| | | AVQZ | 0.19 | 0.06 | 0.08 | 0.35 | 0.14 | 0.26 |
| | | AV5Z | — | 0.02 | — | — | — | — |
| | | AV{T,Q} | 0.25 | 0.33 | 0.39 | 0.10 | 0.22 | 0.30 |
| | | AV{Q,5} | — | 0.05 | — | — | — | — |
| | VeryTightPNO | AVTZ | 0.56 | 0.36 | 0.50 | 0.72 | 0.02 | 0.65 |
| | | AVQZ | — | 0.01 | — | — | — | — |
| | | AV{T,Q} | — | 0.25 | — | — | — | — |
| | NormalPNO | AVTZ | 0.74 | 0.70 | 0.91 | 0.63 | -0.26 | 0.81 |
| | | AVQZ | 0.30 | 0.35 | 0.42 | 0.18 | -0.18 | 0.34 |
| | | AV5Z | 0.24 | 0.30 | 0.36 | 0.11 | -0.21 | 0.28 |
| | | AV{T,Q}Z | 0.13 | 0.11 | 0.10 | 0.17 | -0.13 | 0.15 |
| | | AV{Q,5}Z | 0.24 | 0.25 | 0.30 | 0.18 | -0.23 | 0.25 |
| | TightPNO $T_{CutPNO}=10^{-6}$ | AVTZ | 0.54 | 0.32 | 0.46 | 0.72 | 0.04 | 0.63 |
| | | AVQZ | 0.11 | 0.05 | 0.07 | 0.18 | 0.09 | 0.13 |
| | | AV5Z | — | 0.13 | — | — | — | — |
| | | AV{T,Q}Z | 0.28 | 0.29 | 0.37 | 0.20 | 0.13 | 0.31 |
| | | AV{Q,5}Z | — | 0.22 | — | — | — | — |
| DLPNO-CCSD($T_1$) | TightPNO (Default or $T_{CutPNO}=10^{-7}$) | AVTZ | 0.65 | 0.41 | 0.58 | 0.85 | 0.02 | 0.76 |
| | | AVQZ | 0.15 | 0.05 | 0.07 | 0.28 | 0.07 | 0.21 |
| | | AV5Z | — | 0.03 | — | — | — | — |
| | | AV{T,Q}Z | 0.18 | 0.20 | 0.27 | 0.10 | 0.11 | 0.21 |
| | | AV{Q,5}Z | — | 0.08 | — | — | — | — |
| | TightPNO $T_{CutPNO}=10^{-\{6,7\}}$ or CPS{6,7}[a] | AVTZ | 0.71 | 0.45 | 0.64 | 0.92 | 0.02 | 0.83 |
| | | AVQZ | 0.20 | 0.08 | 0.12 | 0.34 | 0.07 | 0.26 |
| | | AV5Z | — | 0.04 | — | — | — | — |
| | | AV{T,Q}Z | 0.14 | 0.16 | 0.22 | 0.06 | 0.10 | 0.17 |
| | | AV{Q,5}Z | — | 0.02 | — | — | — | — |
| | VeryTightPNO | AVTZ | 0.59 | 0.39 | 0.54 | 0.75 | 0.01 | 0.68 |
| | | AVQZ | — | 0.02 | — | — | — | — |
| | | AV{T,Q}Z | — | 0.23 | — | — | — | — |

[a]The expression CPS{X,Y} refers to the extrapolation of $T_{CutPNO}$ to the complete PNO space limit using $T_{CutPNO}=10^{-X}$ and $10^{-Y}$, where Y=X+1

With the "Default" accuracy threshold and AVTZ basis set, PNO-LCCSD(T) performs remarkably well (0.05 kcal/mol) owing to fortunate error compensation between LO error and basis set incompleteness error. Increasing the basis set size from AVTZ to AVQZ adversely affects the accuracy (MAD increases from 0.05 to 0.11 kcal/mol) as LO error dominates with a larger basis set. With the "Tight" setting, increasing the basis set size from AVTZ to AVQZ has no significant influence on PNO-LCCSD(T) performance when the full ACONFL is considered. Closer scrutiny suggests using a larger basis set degrades the performance of ACONF12 and ACONF16 but marginally improves the accuracy of ACONF20 (see Table 5). Due to the substantial computational cost, we were only able to calculate PNO-LCCSD(T)/AV5Z level conformer energies for ACONF12 and ACONF16 subsets with "Default" and ACONF12 subset with "Tight" cutoffs. Either with the "Default" or "Tight" threshold, PNO-LCCSD(T)/AV5Z and PNO-LCCSD(T)/AVQZ offer comparable performance. Irrespective of the choice of accuracy settings, complete basis set extrapolation adversely affects the performance of standard PNO-LCCSD(T) methods.

Turning to the DLPNO-CCSD($T_1$) with the "NormalPNO" setting, increasing the basis set size improves accuracy as it should (see Table 5). A two-point CBS extrapolation from the AVTZ and AVQZ level conformer energies improves accuracy further (MAD=0.13 kcal/mol). However, a {Q,5} extrapolation only offers accuracy similar to DLPNO-CCSD($T_1$)/AV5Z. With the "TightPNO" threshold and AVQZ basis set, tightening the $T_{CutPNO}$ parameter from $10^{-6}$ to $10^{-7}$ only marginally degrades their performance. Due to the huge I/O, bandwidth, and storage requirements, we were only able to calculate the ACONF12 conformer energies at DLPNO-CCSD($T_1$)/AV5Z and DLPNO-CCSD($T_1$)/AVQZ levels with "TightPNO" and "VeryTightPNO" settings, respectively. With the default "TightPNO" (i.e., $T_{CutPNO}=10^{-7}$) DLPNO-CCSD($T_1$)/AV{T,Q}Z and DLPNO-CCSD($T_1$)/AVQZ offer similar accuracy, which is due to a significant reduction of MAD for ACONF20 while using the prior method (see Table 5). With the default TightPNO and TightPNO at the complete PNO space limit (i.e., $T_{CutPNO}=10^{-\{6,7\}}$), increasing the basis set size from AVQZ to AV5Z marginally improves the accuracy for *n*-dodecane conformers. Using a {T,Q} CBS extrapolation and CPS extrapolation from the $T_{CutPNO}=10^{-6}$ and $10^{-7}$ energies, DLPNO-CCSD($T_1$)/TightPNO can achieve 0.14 kcal/mol accuracy, which is in the territory of LNO-CCSD(T, Tight)/AV5Z and LNO-CCSD(T, vTight)/AVQZ. The CPS{6,7} extrapolation with a two-point CBS extrapolation can significantly improve the performance of the ACONF20 subset.

From the MAD values listed in Table 5, it is clear that for the twelve *n*-dodecane conformers, the performance improvement from "TightPNO" to "VeryTightPNO" is not statistically significant.

Now, if one is limited to the I/O, storage, and bandwidth requirements of the DLPNO-CCSD($T_1$) method, another alternative is to use DLPNO-CCSD($T_0$) instead. At any specific accuracy threshold, the performance of DLPNO-CCSD($T_1$) and DLPNO-CCSD($T_0$) is comparable; this is not surprising, as the conformers of longer *n*-alkanes do not have significant type A static correlation.[77] The only exception to the above trend is TightPNO/AV{T,Q}Z with CPS, where DLPNO-CCSD($T_1$) and DLPNO-CCSD($T_0$) have 0.14 and 0.25 kcal/mol mean absolute deviations, respectively. For the ACONF16 and ACONF12 subsets, DLPNO-CCSD($T_1$, TightPNO)/AV{T,Q}Z/CPS significantly outperforms DLPNO-CCSD($T_0$, TightPNO)/AV{T,Q}Z/CPS (MAD values decrease from 0.33 and 0.39 to 0.16 and 0.22 kcal/mol, respectively).

For organometallic barrier heights, Iron and Janes,[41,42] and later Efremenko and Martin,[39] found that the ($T_1$) – ($T_0$) difference is only weakly sensitive to the basis set size. Hence, in an earlier study on the S66x8 set, we considered a two-tier composite method, DLPNO-CCSD($T_0$)/haVQZ + $c_1$[DLPNO-CCSD($T_0$)/haVQZ – DLPNO-CCSD($T_0$)/haVTZ] + $c_2$[DLPNO-CCSD($T_1$)/haVTZ – DLPNO-CCSD($T_0$)/haVTZ], where the CBS extrapolation is carried out at the DLPNO-CCSD($T_0$) level and the ($T_1$) – ($T_0$) difference is evaluated in a smaller basis set. For the counterpoise uncorrected noncovalent interaction energies, the optimized prefactors were {$c_1$,$c_2$}={0.61, 3.33}. With the dataset in hand, we found that ($T_0$)/AVQZ + 0.61[($T_0$)/AVQZ – ($T_0$)/AVTZ] + 3.33[($T_1$)/AVTZ – ($T_0$)/AVTZ] is only marginally better (MAD=0.10 kcal/mol) than DLPNO-CCSD($T_0$)/AVQZ and DLPNO-CCSD($T_1$)/AVQZ with the "TightPNO" setting (see Table 5 and Table S6 in the Supporting Information). As earlier, the anomalous values of the coefficients seem hard to justify.

Additionally, we also considered benchmarking the LNO, PNO, and DLPNO-based composite methods optimized for the counterpoise uncorrected (or "raw") S66x8 noncovalent interactions (see Ref.[50] for further details). The RMSD, MAD, and MSD statistics of these methods for ACONFL are listed in Table S6. While using the coefficients from Ref.[50], none of the composite schemes outperformed their respective standard methods. On the other hand, optimizing the coefficients of the localized orbital composite methods with respect to the revised ACONFL reference data significantly improved their performance. However, once again the new optimized prefactors seem unphysical sometimes, e.g., for the LNO-based Tight{T,Q} + $c_1$[vTight – Tight]/T method, we got the lowest MAD with $c_1$=-1.47 (see Table S6 in the Supporting Information). Hence, we are reluctant to recommend their use.

As a parenthetical remark, for the RMSD, MAD, and MSD statistics of different localized orbital methods relative to the reference conformer energies using HLC8 instead of HLC14 for the ACONF16 and ACONF20 subsets, see Table S7 in the Supporting Information.

*(d) Explicitly correlated localized orbital coupled cluster methods:*

In this section, we assess the performance of explicitly correlated PNO-LCCSD(T)-F12b and DLPNO-CCSD($T_1$)-F12 methods in combination with different accuracy thresholds and basis sets. The mean absolute deviations (MADs), mean signed deviations (MSDs), and root-mean-square deviations (RMSDs) of these methods are listed in Table 6. Following the recommendation of Peterson *et al.*[75], the (T) components of the PNO-LCCSD(T)-F12b and DLPNO-CCSD($T_1$)-F12 energies are scaled by the global scale factors 1.1413, 1.0527, and 1.0232, respectively, for the VDZ-F12, VTZ-F12, and VQZ-F12 basis sets.

With "Default" cutoffs, PNO-LCCSD(T)-F12b/VDZ-F12 and PNO-LCCSD(T)-F12b/VTZ-F12 offer similar accuracy, which is marginally improved by using (T)-scaling (see Table 6). With the "Tight" setting, using the VTZ-F12 basis set, we obtained marginally better performance than VDZ-F12. However, (T)-scaling closed that gap between PNO-LCCSD(Ts)-F12b/Tight/VDZ-F12 and PNO-LCCSD(Ts)-F12b/Tight/VTZ-F12. PNO-LCCSD(Ts)-F12b/VDZ-F12 offers performance similar to PNO-LCCSD(T)/AVTZ (MAD = 0.09 kcal/mol) when "Tight" cutoffs are used. Interestingly enough, tightening the threshold from "Default" to "Tight" does not offer any noticeable advantage for PNO-LCCSD(Ts)-F12b/VDZ-F12. The "Default" and "Tight" settings yield results of comparable quality for ACONF16 with the VDZ-F12 basis set. However, for the ACONF20 set, we get better performance with a tighter threshold.

Increasing the basis set size from VDZ-F12 to VTZ-F12 does more harm than good for DLPNO-CCSD($T_1$)-F12 when using the "NormalPNO" setting, which is an indication of fortuitous error compensation between errors due to the use of loose accuracy cutoffs and basis set incompleteness. As expected, tightening the accuracy threshold from NormalPNO to TightPNO improves the accuracy of DLPNO-CCSD($T_1$)-F12/VDZ-F12, which gets even better by using the "VeryTightPNO" setting (see Table 6). Increasing the basis set size from VDZ-F12 to VTZ-F12 significantly worsens the performance of DLPNO-CCSD($T_1$)-F12 when the "NormalPNO" threshold is used. However, with "TightPNO" cutoffs, the MAD difference between VDZ-F12 and VTZ-F12 basis sets is statistically insignificant. From the mean absolute deviations listed in Table 6, it is clear that VDZ-F12 reaches the basis set limit with the "TightPNO" setting.

**Table 6:** Performance of explicitly correlated PNO-LCCSD(T)-F12b, PNO-LCCSD(Ts)-F12b, DLPNO-CCSD($T_1$)-F12, and DLPNO-CCSD($T_{1s}$)-F12 methods with respect to the revised ACONFL reference data. Heatmapping is from red (worst) via yellow to green (best).[a]

| Method | Threshold | Basis set | MAD(kcal/mol) | | | | MSD (kcal/mol) | RMSD (kcal/mol) |
|---|---|---|---|---|---|---|---|---|
| | | | ACONFL | ACONF12 | ACONF16 | ACONF20 | | |
| PNO-LCCSD(T)-F12b | Default | VDZ-F12 | 0.18 | 0.12 | 0.16 | 0.24 | -0.02 | 0.21 |
| | | VTZ-F12 | 0.16 | 0.15 | 0.19 | 0.16 | 0.04 | 0.18 |
| | | VQZ-F12 | — | 0.10 | — | — | — | — |
| | Tight | VDZ-F12 | 0.18 | 0.15 | 0.19 | 0.18 | 0.03 | 0.19 |
| | | VTZ-F12 | 0.13 | 0.12 | 0.15 | 0.12 | 0.04 | 0.14 |
| PNO-LCCSD(Ts)-F12b | Default | VDZ-F12 | 0.11 | 0.06 | 0.08 | 0.17 | -0.03 | 0.13 |
| | | VTZ-F12 | 0.13 | 0.12 | 0.15 | 0.12 | 0.04 | 0.14 |
| | | VQZ-F12 | — | 0.09 | — | — | — | — |
| | Tight | VDZ-F12 | 0.10 | 0.08 | 0.11 | 0.09 | 0.02 | 0.11 |
| | | VTZ-F12 | 0.10 | 0.09 | 0.12 | 0.09 | 0.03 | 0.11 |
| DLPNO-CCSD($T_1$)-F12 | NormalPNO | VDZ-F12 | 0.15 | 0.09 | 0.09 | 0.24 | -0.15 | 0.18 |
| | | VTZ-F12 | 0.36 | 0.27 | 0.40 | 0.39 | -0.35 | 0.57 |
| | | VQZ-F12 | — | 0.34 | — | — | — | — |
| | TightPNO | VDZ-F12 | 0.08 | 0.04 | 0.07 | 0.11 | -0.02 | 0.10 |
| | | VTZ-F12 | 0.12 | 0.06 | 0.07 | 0.19 | 0.03 | 0.19 |
| | | VQZ-F12 | — | 0.03 | — | — | — | — |
| | VeryTightPNO | VDZ-F12 | 0.04 | 0.04 | 0.05 | 0.04 | 0.01 | 0.05 |
| | | VTZ-F12 | — | 0.06 | — | — | — | — |
| | | VQZ-F12 | — | 0.07 | — | — | — | — |
| DLPNO-CCSD($T_{1s}$)-F12 | NormalPNO | VDZ-F12 | 0.18 | 0.14 | 0.16 | 0.23 | -0.18 | 0.20 |
| | | VTZ-F12 | 0.38 | 0.29 | 0.43 | 0.39 | -0.37 | 0.58 |
| | | VQZ-F12 | — | 0.35 | — | — | — | — |
| | TightPNO | VDZ-F12 | 0.16 | 0.11 | 0.15 | 0.18 | -0.03 | 0.18 |
| | | VTZ-F12 | 0.12 | 0.04 | 0.06 | 0.20 | 0.03 | 0.18 |
| | | VQZ-F12 | — | 0.02 | — | — | — | — |
| | VeryTightPNO | VDZ-F12 | 0.05 | 0.03 | 0.04 | 0.07 | 0.00 | 0.06 |
| | | VTZ-F12 | — | 0.03 | — | — | — | — |
| | | VQZ-F12 | — | 0.06 | — | — | — | — |

[a] Following Ref.[75], the (T) terms of PNO-LCCSD(T)-F12b and DLPNO-CCSD($T_1$)-F12 were scaled by 1.1413, 1.0527, and 1.0232, respectively, for the VDZ-F12, VTZ-F12, and VQZ-F12 basis sets.

With the "NormalPNO" settings, the performance of explicitly correlated DLPNO-CCSD($T_1$)-F12/VDZ-F12 is very close to standard DLPNO-CCSD($T_1$)/AV{T,Q}Z. With a mean absolute deviation of 0.04 kcal/mol, in fact, DLPNO-CCSD($T_1$)-F12/VeryTightPNO/VDZ-F12 outperforms all the standard DLPNO-CCSD($T_1$) tested in the present study.

*(e) Prototypical timing comparison:*

In this section, we compare how costly different localized coupled cluster methods are relative to DF-CCSD(T). We have considered one *n*-dodecane conformer (to be more specific, conformer **0**). For the standard and explicitly correlated calculations, the aug-cc-pVTZ basis set was employed throughout. All the calculations were carried out on 16 cores of a node with two Intel(R) Xeon(R) Gold 5320 CPUs (2.20GHz).

**Table 7:** Total wall time (hrs.) for an *n*-dodecane conformer with DF-CCSD(T) and different localized coupled cluster methods using different accuracy thresholds. The aug-cc-pVTZ basis set and 16 Intel(R) Xeon(R) Gold 5320 CPU (2.20GHz) cores were used throughout.

| Method | Threshold | Wall Time (hrs.) |
|---|---|---|
| DF-CCSD(T) |  | 74.82 |
| DLPNO-CCSD($T_1$) | NormalPNO | 0.83 |
|  | TightPNO ($T_{CutPNO}=10^{-6}$) | 1.16 |
|  | TightPNO (Default or $T_{CutPNO}=10^{-7}$) | 2.29 |
|  | VeryTightPNO | 9.25 |
| LNO-CCSD(T) | Normal | 0.23 |
|  | Tight | 0.57 |
|  | vTight | 1.69 |
|  | vvTight | 4.66 |
| PNO-CCSD(T) | Default | 0.46 |
|  | Tight | 0.75 |
| PNO-LCCSD(T)-F12b | Default | 0.52 |
|  | Tight | 1.14 |
| DLPNO-CCSD($T_1$)-F12 | NormalPNO | 2.69 |
|  | TightPNO | 4.69 |
|  | VeryTightPNO | 12.58 |

Among the standard localized coupled cluster methods, DLPNO-CCSD($T_1$) with a VeryTightPNO setting (i.e., TightPNO with $T_{CutPNO}=10^{-8}$, $T_{CutMKN}=10^{-4}$, & $T_{CutPairs}=10^{-6}$) is the most expensive one, at 1/8th the cost of DF-CCSD(T). For the TightPNO setting, loosening the $T_{CutPNO}$ one notch from the default value (i.e., $10^{-7}$) reduces the cost of DLPNO-CCSD($T_1$) approximately by half. For LNO-CCSD(T), tightening the accuracy thresholds from Normal to Tight, Tight to vTight, and vTight to vvTight increases the cost by 2.5, 3.0, and 2.8 times respectively. The LNO-CCSD(T) calculations with a vvTight threshold are computationally twice as expensive as standard DLPNO-CCSD($T_1$,TightPNO). LNO-CCSD(T) with the Normal threshold is the least expensive localized coupled cluster method among the methods listed in Table 7. For PNO-LCCSD(T), the total wall time ratio for "Default" and "Tight" thresholds is 1:1.63. With the "Tight" setting, PNO-LCCSD(T) is approximately as expensive as DLPNO-CCSD($T_1$, NormalPNO).

For PNO-LCCSD(T)-F12b, the Default:Tight ratio for total wall time is 1:2.2. With a "Tight" threshold, explicitly correlated PNO-LCCSD(T)-F12b is just 1.5 times more expensive than standard PNO-LCCSD(T).

On the other hand, DLNO-CCSD($T_1$)-F12 calculations are 3.2, 2.0, and 1.4 times more expensive than regular DLPNO-CCSD($T_1$) with "NormalPNO", "TightPNO", and "VeryTightPNO" cutoffs, respectively.

*(f) A few remarks on the performance of more approximate methods:*

In this subsection, we try to answer the question: to what extent do the new reference energies of ACONFL affect the performance of DFT, semiempirical quantum mechanical (SQM), and force field (FF) methods?

In ref.[24], Hansen and coworkers evaluated the performance of a variety of such methods. Hence, to save us time, we have extracted the conformer energies from the Supporting Information of ref.[24], spliced in our new reference data, and compared the statistics of different DFT, SQM, and FF methods (see the Excel workbook in the Supporting Information). On top of that, we have considered a few additional double hybrid functionals, e.g., revDSD-PBEP86-D3BJ,[68] revDSD-PBEP86-D4,[68] ωB97M(2),[78] and dRPA-based fifth rung functionals.[71,72,79] The first three functionals were the best performers in the GMTKN55 (general main-group thermochemistry, kinetics, and noncovalent interactions, 55 problem types[37]) benchmark,[68] while dispersion corrected dRPA-based functionals performed remarkably well for S66 and S66x8 noncovalent interactions.[50,80]

**Table 8:** Performance of different double hybrid functionals with respect to the revised reference data for ACONFL.[a]

| Methods | Basis set | MAD (kcal/mol) | | | | MSD (kcal/mol) | RMSD (kcal/mol) |
|---|---|---|---|---|---|---|---|
| | | ACONFL | ACONF12 | ACONF16 | ACONF20 | | |
| ωB97M-D3BJ | def2-QZVPP | 0.12 | 0.19 | 0.13 | 0.07 | -0.10 | 0.14 |
| B2PLYP-D3BJ | def2-QZVPP | 0.21 | 0.17 | 0.22 | 0.22 | 0.06 | 0.22 |
| B2PLYP-D4 | def2-QZVPP | 0.16 | 0.11 | 0.16 | 0.20 | 0.04 | 0.18 |
| PWPB95-D3BJ | def2-QZVPP | 0.13 | 0.16 | 0.11 | 0.13 | -0.11 | 0.15 |
| PWPB95-D4 | def2-QZVPP | 0.16 | 0.21 | 0.13 | 0.16 | -0.11 | 0.19 |
| DSD-BLYP-D3BJ | def2-QZVPP | 0.27 | 0.19 | 0.25 | 0.34 | 0.00 | 0.32 |
| DSD-BLYP-D4 | def2-QZVPP | 0.22 | 0.19 | 0.24 | 0.22 | 0.08 | 0.24 |
| revDSD-BLYP-D4 | def2-QZVPP | 0.25 | 0.24 | 0.27 | 0.24 | 0.11 | 0.29 |
| HF-D4 | def2-QZVPP | 0.41 | 0.24 | 0.33 | 0.58 | -0.05 | 0.49 |
| revDSD-PBEP86-D4 | def2-QZVPP | 0.26 | 0.18 | 0.24 | 0.32 | 0.01 | 0.30 |
| revDOD-PBEP86-D4 | def2-QZVPP | 0.26 | 0.16 | 0.23 | 0.34 | -0.01 | 0.31 |
| revDSD-PBEP86-D3BJ | def2-QZVPP | 0.19 | 0.16 | 0.20 | 0.19 | 0.04 | 0.21 |
| ωB97M(2) | def2-QZVPP | 0.13 | 0.11 | 0.18 | 0.11 | 0.05 | 0.15 |
| dRPA75 | def2-QZVPP | 0.84 | 0.66 | 0.88 | 0.90 | 0.10 | 0.93 |
| | def2-TZVPP | 0.49 | 0.32 | 0.44 | 0.63 | -0.04 | 0.56 |
| | def2-{T,Q}ZVPP | 1.02 | 0.84 | 1.11 | 1.06 | 0.00 | 1.13 |
| dRPA75-D3BJ | def2-QZVPP | 0.04 | 0.02 | 0.04 | 0.06 | 0.03 | 0.05 |
| | def2-TZVPP | 0.37 | 0.34 | 0.41 | 0.37 | -0.11 | 0.41 |
| | def2-{T,Q}ZVPP | 0.18 | 0.18 | 0.26 | 0.12 | 0.00 | 0.21 |
| dRPA75-D4 | def2-QZVPP | 0.14 | 0.07 | 0.11 | 0.21 | -0.04 | 0.17 |
| | def2-TZVPP | 0.26 | 0.28 | 0.33 | 0.19 | -0.18 | 0.29 |
| | def2-{T,Q}ZVPP | 0.32 | 0.24 | 0.34 | 0.35 | 0.00 | 0.36 |
| DSD-PBEdRPA75-D3BJ | def2-QZVPP | 0.04 | 0.02 | 0.06 | 0.03 | 0.03 | 0.05 |
| | def2-TZVPP | 0.31 | 0.28 | 0.34 | 0.29 | -0.10 | 0.33 |
| | def2-{T,Q}ZVPP | 0.19 | 0.18 | 0.26 | 0.14 | 0.00 | 0.21 |
| DSD-PBEdRPA75-D4 | def2-QZVPP | 0.21 | 0.10 | 0.16 | 0.33 | -0.06 | 0.27 |
| | def2-TZVPP | 0.20 | 0.21 | 0.24 | 0.16 | -0.19 | 0.21 |
| | def2-{T,Q}ZVPP | 0.38 | 0.26 | 0.36 | 0.45 | 0.00 | 0.43 |
| DSD-PBEP86dRPA75-D3BJ | def2-QZVPP | 0.08 | 0.06 | 0.09 | 0.09 | 0.08 | 0.09 |
| | def2-TZVPP | 0.28 | 0.22 | 0.27 | 0.33 | -0.03 | 0.32 |
| | def2-{T,Q}ZVPP | 0.19 | 0.20 | 0.28 | 0.11 | 0.00 | 0.22 |
| DSD-PBEP86dRPA75-D4 | def2-QZVPP | 0.19 | 0.12 | 0.18 | 0.24 | -0.01 | 0.22 |
| | def2-TZVPP | 0.14 | 0.16 | 0.18 | 0.10 | -0.12 | 0.16 |
| | def2-{T,Q}ZVPP | 0.34 | 0.27 | 0.37 | 0.36 | 0.00 | 0.38 |

[a]Everything below the dotted line is calculated by ourselves in the present study. For the statistics of the functionals above the dotted line, we have used the conformer energies reported in the ESI of ref.[24]

The mean absolute error of the widely used molecular mechanics force fields, UFF[81] and MMFF94,[82,83] was marginally reduced from 2.91 and 3.41 to 2.87 and 3.37 kcal/mol, respectively. Considering the fact that the average conformer energy of ACONFL is 4.56 kcal/mol, these errors remain unacceptable. With MAD=0.40 kcal/mol, GFN-FF[84] emerges as the best performer among all the force fields, followed by OpenFF-1.0.0[85] (MAD=0.58 kcal/mol). However, it should be mentioned that the surprisingly accurate performance of simple HF with D4 dispersion correction, 0.14 kcal/mol, is now reduced to merely good one, 0.41 kcal/mol.

Now, among the SQM methods, the performance of the best pick, PM6-DH4,[86] gets marginally better when we employ the revised reference energies (MAD goes down from 0.55 to 0.48 kcal/mol). However, for PM7,[87] this improvement is a bit more prominent.

Finally, we will focus on the performance of DFT methods in light of the presently revised reference data. The mean absolute errors of PBE-D4,[88–90] TPSS-D3BJ,[91,92] and B97M-D4[93,94] decrease from 0.33, 0.42, and 0.46 kcal/mol to 0.10, 0.28, and 0.17 kcal/mol, respectively. On the other hand, for SCAN-D3BJ[95] and r$^2$SCAN-D3BJ,[96–98] the MAD values increase from 0.21 to 0.34 kcal/mol and 0.25 to 0.34 kcal/mol, respectively. With D3(BJ),[92,99] D4,[88,100] or VV10[101] dispersion correction, B3LYP[102] and PBE0[88,103,104] perform significantly better now. With MAD=0.07 kcal/mol, PBE0-VV10 is the best pick among the hybrid functionals. That being said, the mean absolute errors of ωB97M-D3BJ[105] and ωB97M-D4[94] decrease from 0.39 and 0.32 to 0.12 and 0.20 kcal/mol, respectively.

With D4 dispersion correction, the excellent performance of B2PLYP[106] stays intact. However, the mean absolute error of PWPB95[107] with D3BJ and D4 improves from 0.35 each to 0.13 and 0.16 kcal/mol, respectively. Performance of PWPB95-D3BJ, ωB97M(2),[78] and the lower-rung ωB97M-D3BJ functionals are statistically indistinguishable (see Table 8). As we saw for the GMTKN55 benchmark,[68] Head-Gordon's 16-paramter ωB97M(2) marginally outperforms our six-parameter revDSD-PBEP86-D3BJ for longer n-alkane conformer energies.

Let us switch our focus to the dRPA-based double hybrid functionals. As all the systems of the ACONFL set are closed-shell, Kállay's dRPA75[71] and SCS-dRPA75[79] are equivalent. Like Brauer et al.[80] found for the S66x8 noncovalent interactions, adding an empirical dispersion correction significantly improved the performance of (SCS-)dRPA75. (With the def2-QZVPP basis set, the MAD values of dRPA75, dRPA75-D3BJ, and dRPA75-D4 are 0.84, 0.04, and 0.14 kcal/mol, respectively.) The main difference between Kállay's dRPA75 and our DSD-PBEdRPA$_{75}$[72] is that the former functional does not have any semi-local correlation in the final energy expression, but the latter one contains a small percentage (~10%) of PBEc correlation. With D3BJ dispersion corrections, dRPA75 and DSD-PBEdRPA$_{75}$ offer the lowest MADs among all the double hybrid DFT methods tested. With a mean absolute error of 0.04 kcal/mol, these two double hybrids are marginally better than DSD-PBEP86dRPA$_{75}$-D3BJ.[72] For DSD-PBEP86dRPA$_{75}$-D4, due to fortunate error compensation, we obtain a marginally better MAD with def2-TZVPP than using a larger def2-QZVPP basis set. Unlike what we observed for the GMTKN55 benchmark,[72] a complete basis set extrapolation from the def2-TZVPP,[108] and def2-QZVPP[108] conformer energies is detrimental to the performance of the dRRA-based double hybrids (see Table 8).

The errors in DFT calculations can be grouped into two camps: imperfections in the functional itself and errors arising from the self-consistent density evaluated with an imperfect functional. In recent years, Sim, Burke, and coworkers[109] have established the theory of density-corrected density functional theory (DC-DFT) to minimize the second class of errors. They propose a straightforward solution using converged Hartree−Fock densities (HF-DFT) instead of self-consistent ones for the final evaluation of the exchange-correlation (XC) functional. (For more details on DC-DFT, see the review article by Wasserman et al.[110] and a recent paper in questions-and-answers format by Song et al.[111])

**Table 9:** Performance of HF- and KS-DFT functionals for large and medium size *n*-alkane conformers (i.e., ACONFL and ACONF[14] sets).

| Methods | ACONFL (i.e., C$_n$H$_{2n+2}$; n=12, 16, & 20) | | | | | | ACONF (i.e., C$_n$H$_{2n+2}$; n=2-7) | | |
|---|---|---|---|---|---|---|---|---|---|
| | MAD (kcal/mol) | | | | MSD (kcal/mol) | RMSD (kcal/mol) | MAD (kcal/mol) | MSD (kcal/mol) | RMSD (kcal/mol) |
| | Full | ACONF12 | ACONF16 | ACONF20 | | | | | |
| PBE[a] | 2.63 | 1.99 | 2.55 | 3.07 | 0.15 | 2.94 | 0.59 | 0.59 | 0.67 |
| HF-PBE | 3.11 | 2.44 | 3.13 | 3.49 | 0.31 | 3.46 | 0.77 | 0.77 | 0.85 |
| PBE-D4[a] | 0.10 | 0.08 | 0.09 | 0.12 | -0.07 | 0.12 | 0.13 | 0.13 | 0.19 |
| HF-PBE-D4 | 0.07 | 0.08 | 0.07 | 0.08 | -0.03 | 0.09 | 0.07 | -0.05 | 0.09 |
| PBE0[a] | 2.60 | 1.96 | 2.54 | 3.02 | 0.17 | 2.91 | 0.62 | 0.62 | 0.70 |
| HF-PBE0 | 2.88 | 2.21 | 2.87 | 3.27 | 0.26 | 3.21 | 0.72 | 0.72 | 0.81 |
| PBE0-D4[a] | 0.08 | 0.07 | 0.09 | 0.09 | 0.04 | 0.10 | 0.18 | 0.18 | 0.21 |
| HF-PBE0-D4 | 0.12 | 0.07 | 0.11 | 0.16 | 0.03 | 0.14 | 0.04 | 0.04 | 0.04 |

[a]The conformer energies are taken from the ESI of ref.[24]

In a previous study,[112] we found that for the unbranched *n*-alkane conformers (*n*=2-7),[14] using HF densities instead of self-consistent KS densities can significantly improve the performance of pure and hybrid PBE-D4 functionals. However, without dispersion correction, the trend is the opposite. Just for the sake of completeness, with the present dataset in hand, we have tested whether the same is true for the longer *n*-alkane conformers (see Table 9). Without a dispersion correction, we find that self-consistent KS densities are preferred over HF ones. On the other hand, with the D4 dispersion correction, the performance of HF-PBE-D4[112] and HF-PBE0-D4[112] is statistically indistinguishable from PBE-D4 and PBE0-D4, respectively.

### IV.    Conclusions.

We have successfully calculated the "silver" standard conformer energies of the ACONF12 set, which are very close to the (T)-scaled CCSD(T)-(F12*)/VTZ-F12 energies. Relative to the presently revised reference conformer energies of *n*-dodecane, DLPNO-CCSD($T_1$,VeryTightPNO)/CBS (i.e., the reference energies of *n*-dodecane conformers used by Ehlert *et al.*[24]) has 0.25 kcal/mol mean absolute error. Searching for an alternative high-level correction based on localized-orbital coupled cluster methods, we found that [LNO-CCSD(T) − LMP2]/vTight/AV{Q,5}Z (i.e., HLC14) can replace the more expensive [CCSD(F12*) − MP2-F12]/VTZ-F12 and (T)/AV{D,T}Z, without noticeably sacrificing accuracy. Hence, we have used HLC14 to calculate the reference conformer energies of the ACONF16 and ACONF20 subsets. Relative to the canonical DF-CCSD(T) conformer energies, tightening the accuracy threshold of localized coupled cluster methods improves their performance. Using a LNO-based composite method, vTight + 0.5[vTight-Tight] (see Ref.[31]) and  extrapolation of the DLPNO-CCSD($T_1$,TightPNO) conformer energies to the complete PNO space limit from the $T_{CutPNO}=\{10^{-6},10^{-7}\}$ energies improves accuracy significantly.

Finally, from an extensive survey of different pure and composite localized coupled cluster methods for conformational energies of longer n-alkane chains, we can conclude the following:

- Increasing the basis set size and/or tightening the accuracy threshold improves the accuracy of the pure LNO-CCSD(T) methods. With the "Normal" setting, the AV{T,Q}Z extrapolation performs better than the more expensive AV{Q,5}Z, but the trend is the opposite when we use the "Tight" or "vTight' setting. With the "Tight" setting, LNO-CCSD(T)/AV{Q,5}Z is the best performer among the LNO-based methods tested; hence the composite methods based on Tight{Q,5} have no additional advantage over the pure method.
- For a certain threshold, increasing the basis set seize from AVTZ to AVQZ helps improve the performance of DLPNO-CCSD($T_1$) significantly. Two-point CBS extrapolation does more harm than good for DLPNO-CCSD($T_1$, TightPNO) when we use $T_{CutPNO}= 10^{-6}$. With

- a MAD of 0.10 kcal/mol, the low-cost three tier composite scheme, $(T_0)$TightPNO/AVQZ + 0.61[$(T_0)$TightPNO/AVQZ – $(T_0)$TightPNO/AVTZ] + 3.33[$(T_1)$TightPNO/AVTZ – $(T_0)$TightPNO/AVTZ] is only marginally better than standard DLPNO-CCSD($T_0$, TightPNO)/AVQZ and DLPNO-CCSD($T_1$, TightPNO)/AVQZ.

- For a specific accuracy threshold and basis set combination, performance of DLPNO-CCSD($T_0$) and DLPNO-CCSD($T_1$) are comparable to each other, which is not surprising because none of the conformers of ACONFL set has significant type A static correlation.[77]

- Employing our previously proposed[50] composite schemes does not offer any advantage over the standard localized orbital methods.

Even with the VDZ-F12 basis set, results of the explicitly correlated PNO-LCCSD(T)-F12b and DLPNO-CCSD($T_1$)-F12 are pretty impressive. With "Tight" accuracy cutoffs, explicitly correlated PNO-LCCSD(Ts)-F12b/VDZ-F12 offers accuracy comparable to standard PNP-LCCSD(T)/AVTZ. Among all the explicitly correlated localized orbital methods tested, DLPNO-CCSD($T_1$)-F12/VDZ-F12 with "VeryTightPNO" is the best pick (MAD=0.04 kcal/mol), which is better than any standard DLPNO-CCSD($T_1$) considered in the present study.

When analyzing the ΔMAD values between old and new reference data, it should be kept in mind that the MAD between the two reference sets *themselves* is 0.31 kcal/mol. (The latter will also be an upper limit for ΔMAD.) Hence, for methods where the MAD already was several times larger than 0.31 kcal/mol, the choice of reference data will not affect any conclusions as to the suitability of the said methods — but for approaches with an MAD comparable to or smaller than 0.31 kcal/mol, the choice of reference set may upend some conclusions.

For all dispersion-uncorrected DFT functionals tested in the present study, the MAD value uniformly increases by 0.30 kcal/mol — but as they already perform so poorly, this is not an issue. That being said, for HF-D4, MAD increases nontrivially from 0.14 to 0.41 kcal/mol when substituting the present reference data, while r$^2$SCAN-VV10, r$^2$SCANh-VV10,[113] and r2SCAN0-VV10[113] see their MADs double from {0.18,0.18,0.17} to {0.35,0.34,0.34} kcal/mol. On the other hand, for ωB97X-V[114] and ωB97M-V[69], the mean absolute errors are more than halved from 0.47 to 0.19 and from 0.54 to 0.24 kcal/mol, respectively.

With the new reference data, the performance of Head-Gordon's combinatorially optimized, range-separated double hybrid ωB97M(2) and Grimme's PWPB95-D3BJ are statistically indistinguishable (MAD=0.13 kcal/mol for both)—this would not have been the case if the old reference data were used (0.18 and 0.35 kcal/mol, respectively).

The MADs of our revDSD-PBEP86-D4 and revDOD-PBEP86-D4 functionals increase from 0.06 and 0.07 kcal/mol versus the old reference, to (both) 0.26 kcal/mol versus the new reference. With the D3BJ dispersion correction, the dRPA-based double hybrid functionals, DSD-PBEdRPA75-D3(BJ) and dRPA75-D3BJ are the two best-performing double hybrids (both MAD=0.04 kcal/mol) against the new reference set, but this would not have been the case for the old reference (0.29 and 0.32 kcal/mol, respectively).

Using the AVTZ basis set, we found that localized orbital coupled cluster methods are two orders of magnitude cheaper than the density-fitted canonical CCSD(T) for the *n*-dodecane conformers. With the "VeryTightPNO" accuracy setting, DLPNO-CCSD($T_1$) is twice as expensive as LNO-CCSD(T,vvTight). However, for larger systems, one will start witnessing the formation of domains that may make the DLPNO-based methods cheaper than both LNO-CCSD(T) and PNO-CCSD(T).

**Acknowledgments:**

GS acknowledges a doctoral fellowship from the Feinberg Graduate School (WIS). We gratefully acknowledge Prof. H.-J. Werner (Institute for Theoretical Chemistry, Universität Stuttgart) for helpful communications concerning the PNO-LCCSD(T) code in MOLPRO, particularly his advice concerning the distance criterion for PAO domain selection, and his kindly rerunning several sets of calculations to rule out programming errors.

This research was funded by the Israel Science Foundation (grant 1969/20), by the Minerva Foundation (grant 20/05), and by a research grant from the Artificial Intelligence and Smart Materials Research Fund, in Memory of Dr. Uriel Arnon.

**Supporting Information:**

PDF file containing Tables S1-S9 and the full Ref.[70]; four Microsoft Excel workbooks containing absolute and relative energies for (a) different localized orbital coupled cluster approaches; (b) F12 approaches; (c) DFT methods; (a) different HLCs in ACONF12.

The Supporting Information is available free of charge at https://doi.org/10.1021/acs.jpca.xxxxxx.

# Table of Contents Graphic:
(7.8cm by 4.5 cm)

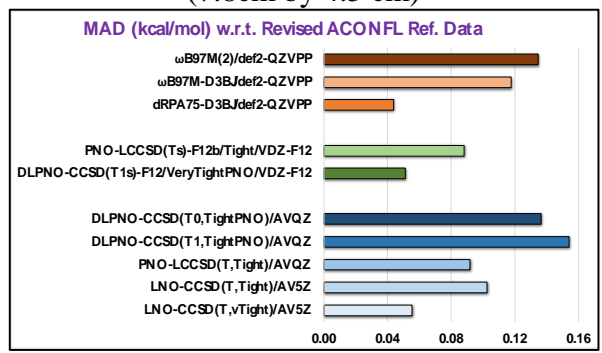